# Path probability of stochastic motion: A functional approach

Masayuki Hattori[a], Sumiyoshi Abe[a,b]

[a] *Department of Physical Engineering, Mie University, Mie 514-8507, Japan*

[b] *Institute of Physics, Kazan Federal University, Kazan 420008, Russia*

ABSTRACT

The path probability of a particle undergoing stochastic motion is studied by the use of functional technique, and the general formula is derived for the path probability distribution functional. The probability of finding paths inside a tube/band, the center of which is stipulated by a given path, is analytically evaluated in a way analogous to continuous measurements in quantum mechanics. Then, the formalism developed here is applied to the stochastic dynamics of stock price in finance.

## 1. Introduction

Consider a certain stochastic equation for $X(t)$ that contains an external noise $\eta(t)$. A solution of the equation satisfying a certain initial condition, $X_\eta(t)$, defines a trajectory of a particle/walker. Then, it is essential to determine the probability of finding the particle in the interval $[x, x+dx]$ at time $t$: $f(x,t)\,dx$, where $f(x,t)$ is the probability distribution function given by $f(x,t) = \langle \delta(x - X_\eta(t)) \rangle_\eta$ with $\langle \bullet \rangle_\eta$ being the average over the noise (see the next section). The passage from the stochastic equation to the probability distribution function is commonly established through the Fokker-Planck equation [1,2]. This is a widely discussed issue that enables one to describe important phenomena such as diffusion and transport.

A less discussed issue may be *path probability*, $P[x]$. In this case, one is concerned with the probability distribution functional that the particle follows the path $x(t)$. One of the first works relevant to this problem may be that in Ref. [3], where irreversible relaxation processes to equilibria are studied for macroscopic thermodynamic variables. Recently, the problem of path probabilities has been revisited in Ref. [4]. It seems, however, that still some points regarding the action functional remain to be clarified in that work.

Our purpose here is to develop the theory of path probabilities in stochastic processes based on the functional method. First, we derive the general formula for the path probability distribution functional associated with the Langevin equation. Then, we examine the overdamped limit, in which the formula becomes simplified drastically. As a byproduct, at this stage, the statement about the action functional made in Ref. [4] is clarified and modified. To obtain a finite measure, we further consider the probability of finding a path inside a "tube", or a "band" in a single dimension, the center of which is prescribed by a given path. In particular, there we employ the Gaussian filtering method inspired by the concept of continuous measurements in quantum mechanics [5]. We discuss these issues in the application of the formula to the stochastic dynamics of stock price in finance.

## 2. Path probability

The path probability distribution functional is defined as follows:

$$P[x] = \left\langle \delta\left[x - X_\eta\right] \right\rangle_\eta, \tag{1}$$

where $\delta\left[x - X_\eta\right]$ is the delta functional given by

$$\delta\left[x - X_\eta\right] = \prod_t \delta\left(x(t) - X_\eta(t)\right) \tag{2}$$

with the symbol $\prod_t$ being the continuous infinite product. The normalization condition is expressed by the functional integral: $\int \mathcal{D}x \, P[x] = 1$, where $\mathcal{D}x$ is the functional integration measure $\mathcal{D}x \equiv \prod_t dx(t)$.

Suppose that $X_\eta(t)$ is a solution of the Langevin equation of a particle with unit mass

$$\frac{d^2 X(t)}{dt^2} = -\kappa \frac{dX(t)}{dt} + F(X(t)) + \sqrt{2D}\,\eta(t) \tag{3}$$

defined in a time interval, $0 < t < T$. Here, $\kappa$ and $D$ are positive constants referred to as the friction and diffusion coefficients, respectively. $F(X(t))$ is a deterministic force term, whereas $\eta(t)$ is the unbiased Gaussian white noise satisfying

$$\langle \eta(t) \rangle_\eta = 0, \qquad \langle \eta(t)\,\eta(t') \rangle_\eta = \delta(t-t'). \tag{4}$$

The diffusion coefficient is pulled out from the noise as in Eq. (3) for later convenience. Einstein's relation states that, in equilibrium, $\kappa D$ is Boltzmann's constant times the temperature of the environment, an influence of which is represented by the noise. The explicit form of the distribution of $\eta(t)$ is given by

$$p[\eta] = \mathcal{N} \exp\left[-\frac{1}{2}\int dt\,\eta^2(t)\right], \tag{5}$$

where $\mathcal{N}$ is the normalization factor and this symbol will commonly be used in the subsequent discussion. The average of a certain quantity, $Q(\eta)$, over the noise is expressed as $\langle Q(\eta) \rangle_\eta = \int \mathcal{D}\eta\, Q(\eta)\, p[\eta]$, and with this Eq. (4) can immediately be ascertained. Actually, the normalization factor is formally divergent. There are some methods to regularize it. One possible way is to discretize the time parameter. This issue will be discussed later in an explicit example (see Section 5).

Let us calculate the path probability in Eq. (1) for a solution $X_\eta(t)$ of Eq. (3). For this purpose, we rewrite Eq. (3) as

$$\Phi(X) \equiv \frac{d^2 X(t)}{dt^2} + \kappa \frac{dX(t)}{dt} - F(X(t)) - \sqrt{2D}\,\eta(t) = 0. \tag{6}$$

Then, holds the following equality:

$$\delta[\Phi(x)] = \left| \mathrm{Det}\left[\frac{\delta \Phi}{\delta x}\right] \right|^{-1} \delta[x - X_\eta]. \tag{7}$$

The factor in front of the delta functional on the right-hand side denotes the functional determinant given by

$$\mathrm{Det}\left[\frac{\delta \Phi(x(t))}{\delta x(t')}\right] = \mathrm{Det}_{t,t'}\left\{\left[\frac{d^2}{dt^2} + \kappa \frac{d}{dt} - F'(x(t))\right]\delta(t-t')\right\}, \tag{8}$$

where $F'(x) = dF(x)/dx$. It is noted that the value of this quantity is not fixed until the temporal boundary condition is imposed on the solution of the Langevin equation. However, we do not specify such a condition explicitly here.

Substituting Eq. (7) into Eq. (1), we obtain the following general formula:

$$P[x] = \mathcal{N} \left| \mathrm{Det}\left[\frac{\delta \Phi}{\delta x}\right] \right| \exp\left\{ -\frac{1}{4D}\int_0^T dt\,[\ddot{x}(t) + \kappa \dot{x}(t) - F(x(t))]^2 \right\}, \tag{9}$$

where the overdots mean the differentiations with respect to time. This shows how the quantity in the exponential is radically different from the ordinary action functional in the Euclidean path integral approach to quantum mechanics, in contrast to the claim made in Ref. [4].

3. **Overdamping**

In the particular case of the overdamped limit, which is realized in many important systems, the inertial term on the left-hand side in Eq. (3) can be ignored, making Eq. (6) to be

$$\Phi(X) \equiv \kappa \frac{dX(t)}{dt} - F(X(t)) - \sqrt{2D}\,\eta(t) = 0. \tag{10}$$

Accordingly, the path probability reads

$$P[x] = \mathcal{N} \left| \mathrm{Det}_{t,t'} \left[ \left( \kappa \frac{d}{dt} - F'(x(t)) \right) \delta(t-t') \right] \right|$$

$$\times \exp\left\{ -\frac{1}{4D} \int_0^T dt \left[ \kappa \dot{x}(t) - F(x(t)) \right]^2 \right\}. \tag{11}$$

Henceforth, $\kappa$ is set equal to unity for the sake of simplicity. The factor in the exponential might look like an action functional (the "seemingly Lagrangian action" postulated in Ref. [4]), but such an interpretation does not seem to be appropriate, since its "kinetic-like term" comes from the friction.

The functional determinant in Eq. (11) turns out to have a simple closed form [6]. First, we recall the formula: $|\text{Det} M| = \exp(\text{Tr} \ln M)$ with

$$M(t,t') = \left[\frac{d}{dt} - F'(x(t))\right]\delta(t-t'). \tag{12}$$

Then, we write $M(t,t') = (d/dt) K(t,t')$ with

$$K(t,t') = \delta(t-t') - F'(x(t'))G(t-t'), \tag{13}$$

where $G$ is Green's function satisfying

$$\frac{d}{dt}G(t-t') = \delta(t-t'). \tag{14}$$

The general solution of Eq. (14) is given by $G(t-t') = \lambda\theta(t-t') - (1-\lambda)\theta(t'-t)$, where $\lambda$ is a constant and $\theta(t)$ is the Heaviside step function: $\theta(t) = 0, \ 1/2, \ 1$ for $t < 0, \ t = 0, \ t > 0$, respectively. Its Fourier integral representation is: $G(t) = \int d\omega \, \tilde{G}(\omega)\exp(i\omega t)$. As the temporal boundary condition, we impose that $\tilde{G}(\omega)$ is analytic in the upper half of the complex-$\omega$ plane in analogy with the Kramers-Kronig dispersion relation that guarantees causality. Then, the special solution reads $\lambda = 0$, leading to

$$G(t-t') = -\theta(t'-t). \tag{15}$$

Absorbing the constant, $\exp[\text{Tr}\ln(d/dt)]$, into the normalization factor and expanding

the logarithm, we find

$$\ln(\text{Det } K) = -\theta(0) \int_0^T dt\, F'(x(t))$$
$$-\frac{1}{2}\int_0^T dt_1 \int_0^T dt_2\, \theta(t_1 - t_2)\theta(t_2 - t_1) F'(x(t_1)) F'(x(t_2)) + \cdots. \quad (16)$$

On the right-hand side, only the first term turns out to survive, giving rise to

$$\ln(\text{Det } K) = -\frac{1}{2}\int_0^T dt\, F'(x(t)). \quad (17)$$

Therefore, we have

$$P[x] = \mathcal{N} \exp\left[-\int_0^T dt\, L(x, \dot{x})\right], \quad (18)$$

where

$$L(x, \dot{x}) = \frac{1}{4D}\left[\dot{x}(t) - F(x(t))\right]^2 + \frac{1}{2}F'(x(t))$$

$$= \frac{1}{4D}\dot{x}^2(t) - V(x(t)) + \frac{d}{dt}\phi(x(t)), \quad (19)$$

where $V = -(1/2)\left[F' + F^2/(2D)\right]$ and $\phi = -[1/(2D)]\int^{x(t)} dx'\, F(x')$. $L$ in Eq. (19) might look like a Lagrangian, but once again it is noted that $\dot{x}(t)$ comes from the friction force.

## 4. Path probability inside a tube along a given path: analog of continuous quantum measurements

Let us study *in the overdamped limit* the probability of finding a path within a tube, i.e., a band, with a given typical width. In particular, we discuss this problem for the Ornstein-Uhlenbeck-like process [1], in which $F$ in Eq. (10) is given by

$$F(x) = -kx, \tag{20}$$

where $k$ is a constant. In this case, the functional determinant in Eq. (11) becomes constant independent of $x$ and therefore can be absorbed into the normalization factor. Consequently, the path probability reads

$$P[x] = \mathcal{N} \exp\left\{ -\frac{1}{4D} \int_0^T dt \left[ \dot{x}(t) + k\,x(t) \right]^2 \right\}. \tag{21}$$

Now, it is of interest to consider the probability of finding paths within a band, the center of which is given by a certain path, $x_*(t)$. To avoid extra complication, here let us consider the case when the width of the band is *kept unchanged along this path*. Such a probability may be calculated by the use of a filtering functional, $w_\Omega$. For the sake of simplicity, here we employ the Gaussian filter:

$$w_\Omega[x; x_*] = \exp\left\{ -\frac{1}{2\Omega^2} \int_0^T dt \left[ x(t) - x_*(t) \right]^2 \right\}, \tag{22}$$

where $\Omega$ stands for the fixed typical width of the *fuzzy* band. Then, we define the probability of finding paths within the band along $x_*(t)$ as follows:

$$P_\Omega \equiv \int \mathcal{D} x \, P[x] \, w_\Omega[x; x_*] , \qquad (23)$$

where the functional integration is performed over all possible paths defined in the interval $[0, T]$.

In the limit $\Omega \to \infty$, $w_\Omega$ converges to unity, and therefore the filter becomes removed. Accordingly, we have

$$\lim_{\Omega \to \infty} P_\Omega = 1 . \qquad (24)$$

We mention that the above manipulation is analogous to that in continuous quantum measurements [5].

## 5. Application to dynamics of stock price

Now, we apply the theory developed above to the stochastic dynamics of stock price. Let $S(t)$ be the stock price of a company at time $t$. It satisfies the following stochastic equation:

$$dS = \mu \, S \, dt + \sigma \, S \, dW , \qquad (25)$$

which is widely used in connection with the Black-Scholes theory for portfolio [2]. $\mu$ is the rate of return, whereas $\sigma$ is the volatility corresponding to $\sqrt{2D}$ in Eq. (10).

Both $\mu$ and $\sigma$ are assumed to be constant for the sake of simplicity. $dW$ is the Wiener process satisfying the Itô rule

$$(dW)^2 = dt. \tag{26}$$

$dW$ corresponds to $\eta dt$, where $\eta$ is the noise discussed earlier. Eq. (25) describes a multiplicative process because of the presence of the product of $S$ and $dW$, which makes it consistent with the positive nature of $S$. It is useful to perform the change of the variable as

$$X = \ln S. \tag{27}$$

Then, the realization of $X$ can take any real value. Noting Eq. (26), the equation for $X$ is found to be

$$dX = \left(\mu - \frac{\sigma^2}{2}\right) dt + \sigma\, dW, \tag{28}$$

which now describes an additive process. This equation can be recast into

$$\Phi(X) = \frac{dX(t)}{dt} - F_0 - \sigma\, \eta(t) = 0, \tag{29}$$

where $F_0 \equiv \mu - \sigma^2/2$ is a constant. Thus, Eq. (10) (with $\kappa \equiv 1$) is further simplified in this model, and the path probability distribution functional is given by

$$P[x] = \mathscr{N} \exp\left\{-\frac{1}{2\sigma^2}\int_0^T dt\left[\dot{x}(t) - F_0\right]^2\right\}. \tag{30}$$

Now, we wish to evaluate the quantity in Eq. (23) for the present system. A problem here is the fact that due to continuity of time the normalization factor $\mathscr{N}$, e.g. in Eq. (30), is formally divergent. To obtain a finite value, we perform discretization of time. In what follows, we discuss this point in detail.

We divide the interval $[0, T]$ and define the discrete time $\{0 = t_0, t_1, t_2, ..., t_N = T\}$, where $t_n = n\,\Delta t$ ($n = 0, 1, 2, ..., N$) with $\Delta t = T/N$. Then, we reexpress Eq. (30) in the following form:

$$P[x] = \mathscr{N} \exp\left\{-\frac{1}{2\tilde{\sigma}^2}\sum_{n=1}^{N}\left(\frac{x_n - x_{n-1}}{\Delta t} - F_0\right)^2\right\}, \tag{31}$$

where $x_n \equiv x(t_n) = x(n\,\Delta t)$ and $\tilde{\sigma}^2 = \sigma^2/\Delta t$. The continuous limit implies $N \to \infty$ and $\tilde{\sigma}^2 \to \infty$ with $\sigma^2$ being kept fixed. Similarly, the filter in Eq. (22) becomes

$$w_\Omega[x; x_*] = \exp\left\{-\frac{1}{2\tilde{\Omega}^2}\sum_{n=1}^{N}\left(x_n - x_{*n}\right)^2\right\}, \tag{32}$$

where $\tilde{\Omega}^2 = \Omega^2/\Delta t$. The $n = 0$ term does not have to be included since the initial condition is usually specified.

Therefore, the probability of finding paths inside the fuzzy band is given by

$$P_\Omega = \mathscr{N}(\alpha)\int_{-\infty}^{\infty}\prod_{m=1}^{N} dx_m$$

$$\times \prod_{n=1}^{N} \exp\left[-\alpha\left(x_n - x_{n-1} - F_0 \Delta t\right)^2 - \beta\left(x_n - x_{*n}\right)^2\right]. \qquad (33)$$

Here, $\prod_{m=1}^{N} dx_m$ is the discretization of the functional integration measure $\mathcal{D}x$. $\alpha$ and $\beta$ are given by $\alpha = 1/\left[2\tilde{\sigma}^2(\Delta t)^2\right]$ and $\beta = 1/\left(2\tilde{\Omega}^2\right)$, respectively. It is noted that the normalization factor depends only on $\alpha$. The multiple Gaussian integrals can be performed, yielding

$$P_\Omega = \mathcal{N}(\alpha)\sqrt{\frac{\pi^N}{\det A^{(N)}(\alpha, \beta)}}\, e^{-S}, \qquad (34)$$

where

$$S = \frac{1}{\det A^{(N)}(\alpha, \beta)} \left\{ \beta \sum_{n=1}^{N} \alpha^n \left[\det A^{(N-n)}(\alpha, \beta)\right]\left(x_{*n} - x_0 - F_0 n \Delta t\right)^2 \right.$$

$$+ \beta^2 \sum_{\substack{m,n=1 \\ (m<n)}}^{N} \alpha^{n-m} \left[\det A^{(N-n)}(\alpha, \beta)\right]\left[\det B^{(m-1)}(\alpha, \beta)\right]$$

$$\left. \times \left[x_{*n} - x_{*m} - (n-m)F_0 \Delta t\right]^2 \right\}. \qquad (35)$$

In these equations, $A^{(n)}(\alpha, \beta)$ and $B^{(n)}(\alpha, \beta)$ are the $n \times n$ symmetric matrices given by

$$A^{(n)}(\alpha, \beta) = \begin{pmatrix} 2\alpha+\beta & -\alpha & 0 & \cdots & & 0 \\ -\alpha & 2\alpha+\beta & -\alpha & & & \vdots \\ 0 & -\alpha & \ddots & & & 0 \\ \vdots & & & & 2\alpha+\beta & -\alpha \\ 0 & \cdots & 0 & & -\alpha & \alpha+\beta \end{pmatrix}, \qquad (36)$$

$$B^{(n)}(\alpha, \beta) = \begin{pmatrix} 2\alpha+\beta & -\alpha & 0 & \cdots & & 0 \\ -\alpha & 2\alpha+\beta & -\alpha & & & \vdots \\ 0 & -\alpha & \ddots & & & 0 \\ \vdots & & & & 2\alpha+\beta & -\alpha \\ 0 & \cdots & 0 & & -\alpha & 2\alpha+\beta \end{pmatrix}, \qquad (37)$$

respectively. In particular, $\det A^{(0)}(\alpha, \beta) = \det B^{(0)}(\alpha, \beta) \equiv 1$, $\det A^{(1)}(\alpha, \beta) = \alpha + \beta$ and $\det B^{(1)}(\alpha, \beta) = 2\alpha + \beta$ are understood in Eq. (35). The one and only difference between the matrices in Eqs. (36) and (37) is in their $(n, n)$ elements. Recalling that the limit $\Omega \to \infty$ corresponds to $\beta \to 0$, we find the normalization condition in Eq. (24) to yield $\mathcal{N}(\alpha) = \sqrt{\det A^{(N)}(\alpha, 0)/\pi^N} = (\alpha/\pi)^{N/2}$. Thus, we obtain the following expression for the probability:

$$P_\Omega = \sqrt{\frac{\alpha^N}{\det A^{(N)}(\alpha, \beta)}} \, e^{-S}. \qquad (38)$$

Finally, let us further evaluate Eq. (38) for a couple of simple examples of the path $x_*$: (i) the most probable path and (ii) the constant path.

(i) The most probable path. As can be seen in Eq. (30), this path is defined by the equation $\dot{x}_*(t) = F_0$, the discretization of which is $x_{*n} - x_{*n-1} = F_0 \Delta t$. The solution satisfying the initial condition $x_{*0} = x_0$ is $x_{*n} = x_0 + F_0 n \Delta t$.

(ii) The constant path. In this case, the solution remains as the initial condition $x_{*n} = x_0$.

In Fig. 1, we present plots of $P_\Omega$ with respect to the elapse of time $T$ in the two cases (i) and (ii) mentioned above. As expected, the probability associated with the most probable path is larger than that for the constant path (except the initial value taken commonly in both cases). Fig. 2 shows that the probability decays as the exponential law. These results may allow one to estimate the value of future stock price with finite uncertainties.

## 6. Concluding remarks

We have developed the discussion about the path probability of a particle/walker undergoing stochastic motion based on the functional approach. We have derived the general formula for the path probability distribution functional and have evaluated its overdamped limit. Through these, we have clarified the problem regarding the action functional addressed in Ref. [4]. Then, we have considered the probability of finding paths inside a fuzzy tube/band, the center of which is given by a certain path. This has been performed in a way analogous to continuous quantum measurements. We have applied this theoretical framework to the stochastic dynamics of stock price in finance.

Discretizing the time parameter, we have obtained the finite regularized path probability and have analyzed how it changes in terms of the period of time.

Tracking stochastic trajectories is of current interest. Recent developments such as the technique of single-molecule tracking are expected to explore reaction pathways in living cells (see Ref. [7] and the references therein). In finance, the discussion such as that in Section 5 may be relevant to the concept of "range forward contracts" [8]. The theory of path probabilities developed here may play important roles in these areas.

Throughout the present work, we have assumed the Gaussian white noise and Wiener process. A possible generalization to the case of colored noises and non-Markovian processes may also be of obvious interest. If the delta function in Eq. (4) is replaced by a function of finite width, then it is necessary to introduce a corresponding memory kernel to the friction term in the Langevin equation, in general [9]. Non-Markovianity is expected to be crucial for understanding the nature of dynamics of market just after its crashes, where stock price obeys a law that extremely resembles the Omori law for earthquake aftershocks [10]. And, it is known for earthquake aftershocks [11-13] that the associated point processes are non-Markovian and exhibit the aging phenomenon. These issues are further to be clarified.

**Acknowledgments**

One of the authors (S.A.) has been supported in part by a Grant-in-Aid for Scientific Research from the Japan Society for the Promotion of Science (No. 26400391) and by the Ministry of Education and Science of the Russian Federation (the

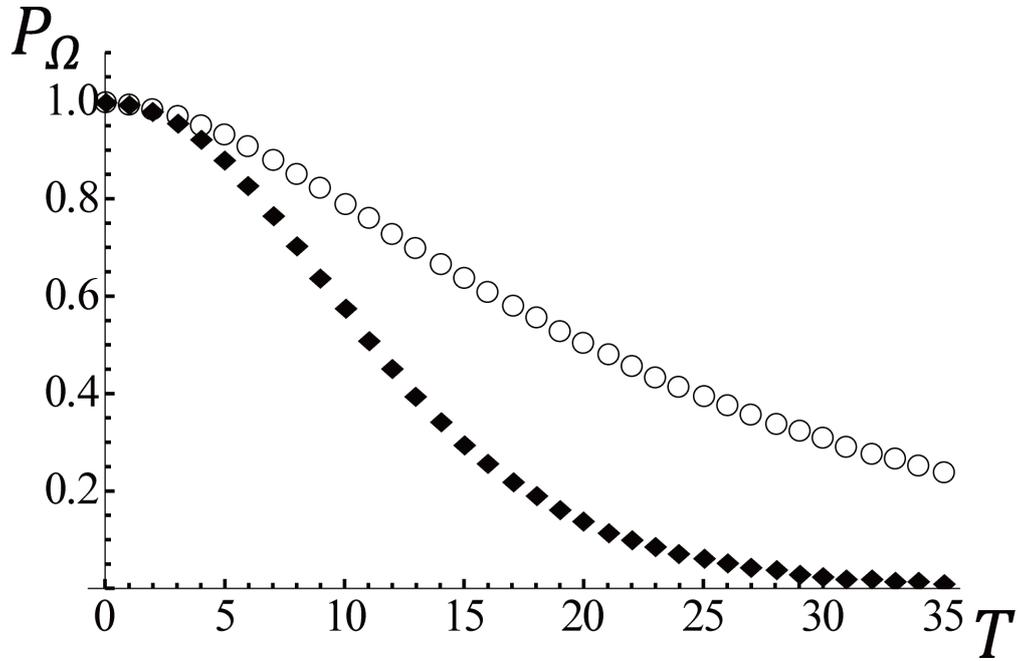

**Fig. 1.** Linear plots of the dimensionless probability $P_\Omega$ with respect to elapse of time $T$ (days). The open circles and filled squares are for (i) the most probable path and (ii) the constant path used for the filters, respectively. The values of the parameters employed are $\sigma^2 = 0.04 \times 10^{-2}$ $(\text{day})^{-1}$, $\mu = 0.01$ $(\text{day})^{-1}$, $\Delta t = 1$ day, $\Omega^2 = 0.04$ day and $x_0 = 1$.

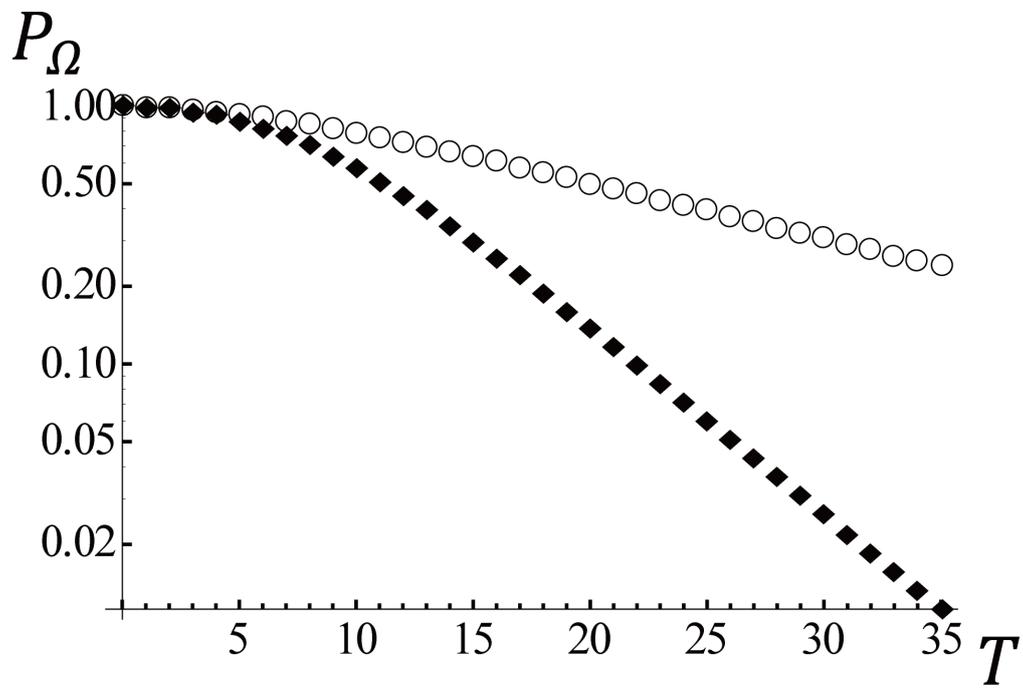

**Fig. 2.** Semi-log plots of the results in Fig. 1.